\documentclass[aps,prb,twocolumn,showpacs,preprintnumbers,amsmath,amssymb,longbibliography]{revtex4-1}

\usepackage{graphicx}
\usepackage[utf8]{inputenc}
\usepackage[pdftex,colorlinks=true,pdfstartview=FitV,linkcolor=blue,citecolor=blue,urlcolor=blue]{hyperref}

\begin{document}

\title{A deep learning approach for determining the chiral indices of carbon nanotubes from high-resolution transmission electron microscopy images}

\author{G. D. Förster}
\affiliation{Laboratoire d'Etude des Microstructures, ONERA-CNRS, UMR104, Université Paris-Saclay, BP 72, 92322 Châtillon Cedex, France}
\affiliation{Aix Marseille Univ, CNRS, CINaM, Marseille, France}
\author{A. Castan}
\affiliation{Laboratoire d'Etude des Microstructures, ONERA-CNRS, UMR104, Université Paris-Saclay, BP 72, 92322 Châtillon Cedex, France}
\affiliation{Department of Physics and Astronomy, University of Pennsylvania, Philadelphia, PA 19104, USA}
\author{A. Loiseau}
\affiliation{Laboratoire d'Etude des Microstructures, ONERA-CNRS, UMR104, Université Paris-Saclay, BP 72, 92322 Châtillon Cedex, France}
\author{J. Nelayah}
\affiliation{Université de Paris, Laboratoire Matériaux et Phénomènes Quantiques, CNRS, F-75013, Paris, France}
\author{D. Alloyeau}
\affiliation{Université de Paris, Laboratoire Matériaux et Phénomènes Quantiques, CNRS, F-75013, Paris, France}
\author{F. Fossard}
\affiliation{Laboratoire d'Etude des Microstructures, ONERA-CNRS, UMR104, Université Paris-Saclay, BP 72, 92322 Châtillon Cedex, France}
\author{C. Bichara}
\affiliation{Aix Marseille Univ, CNRS, CINaM, Marseille, France}
\author{H. Amara}
\affiliation{Laboratoire d'Etude des Microstructures, ONERA-CNRS, UMR104, Université Paris-Saclay, BP 72, 92322 Châtillon Cedex, France}
\affiliation{Université de Paris, Laboratoire Matériaux et Phénomènes Quantiques, CNRS, F-75013, Paris, France}

\email{daniel.foerster2@gmail.com}

\date{\today}

\begin{abstract}
Chiral indices determine important properties of carbon nanotubes (CNTs). Unfortunately, their determination from high-resolution transmission electron microscopy (HRTEM) images, the most accurate method for assigning chirality, is a tedious task. We develop a Convolutional Neural Network that automatizes this process. A large and realistic training data set of CNT images is obtained by means of atomistic computer simulations coupled with the multi-slice approach for image generation. In most cases, results of the automated assignment are in excellent agreement with manual classification, and the origin of failures is identified. The current approach, which combines HRTEM imaging and deep learning algorithms allows the analysis of a statistically significant number of HRTEM images of carbon nanotubes, paving the way for robust estimates of experimental chiral distributions.
\end{abstract}

\keywords{Carbon nanotubes, chiral indices, HRTEM, deep-learning, image classification}

\maketitle

\section{Introduction}

Since their discovery in 1991~\cite{Iijima1991}, single-walled carbon nanotubes (CNTs) are a model system in nanoscience and have attracted tremendous research effort. One of the fundamental reasons is the particular correlation between their structure and their exceptional electrical, thermal, optical and mechanical properties~\cite{Popov2004,Scarselli2012,Rajter2013}. The so-called chiral indices $(n,m)$ entirely determine their atomic structure and characterize their semiconducting or metallic properties~\cite{Hamada1992}. Alternatively, the structure of the CNT can also be described unambiguously by the diameter of the tube and its chiral angle, which is the angle at which the graphene sheet is wrapped to form a nanotube. Therefore, a critical issue is the robust $(n,m)$ identification for individual CNTs to perform statistical analysis of experimental samples. This is particularly important for potential applications such as transparent conductive films~\cite{Ding2017,Tonkikh2019} and transistors~\cite{Tans1998,Franklin2012}. Recently, electronic circuits with hundreds of CNT-based transistors~\cite{Shulaker2013} and even general purpose CPUs functioning solely with CNT transistors have been produced~\cite{Hills2019}.

Experimental techniques for structure determination are based on three routes: spectroscopy, electron diffraction, and HRTEM methods. Firstly, spectroscopic methods (Raman, optical absorption, photo-luminescence excitation) allow for fast and high throughput analysis, but often depend on the environment and cross-sections may vary significantly between chiralities~\cite{Vialla2013,Kataura1999,Lefebvre2004,Dresselhaus2005,Meyer2005,Wang2005,Oyama2006,Castan2018,sanchez2016n}. Secondly, a full identification of $(n,m)$ indices can be extracted from the electron diffraction (ED) patterns obtained with a transmission electron microscope (TEM)~\cite{Colomer2001, Kociak2002, kociak2003accurate, Zuo2003, Zhu2005, Jiang2007, Aleman2016, Zhang2016, Ding2017, He2018b}. While, in principle, ED patterns contain the most precise information on the layer lines (large $q$-vectors) and thus chiral angles of the CNTs, ED is not always feasible because it requires specific apparatus as well as CNT samples showing long sections of clean and well separated nanotubes. Thirdly, the structure of CNTs can also be determined by using phase contrast high-resolution imaging (the so-called HRTEM technique) thanks to recent developments of aberration correctors delivering a resolution in the sub-Ångström range with single atom sensitivity~\cite{Haider1998, Batson2002, Hirahara2006, Alloyeau2012}. As a result, the direct identification of the $(n, m)$ chiral indices of CNTs from atomically resolved images is tractable~\cite{Warner2011, Ghedjatti2017}. Indeed, the moiré pattern emerging from the superposition of the back and front walls of the CNT, with respect to the direction of the incident electron beam, can be compared with reference images obtained from simulated HRTEM images. With recent advances in microcopy techniques, it has become increasingly common to obtain large amounts of HRTEM data. However, the process of determining the chiral indices using this approach is tedious and time consuming if performed manually as it is the case today. Studies based on this method can thus afford only statistics on a small set of CNTs visible in the sample. With the present contribution focused on the HRTEM route to chirality determination, we aim at the automatization of this process enabling thereby the analysis of a statistically significant number of HRTEM images of CNTs.

Recently, deep learning, and in particular convolutional neural networks (CNNs), have shown outstanding performance in visual classification tasks. Since the first pioneering work in this field~\cite{Lecun1998}, the technique has been refined and is now readily applied on many challenging problems~\cite{Carleo2019}, such as the classification of complex objects within photos in one of thousands of categories~\cite{Touvron2019}. The task of chirality assignment on the basis of HRTEM images of CNTs is rather 
simple and comparable to the first successful applications of CNNs: Grayscale images of simple geometrical objects need to be classified into one of a relatively small number of possible classes. Specifically, the convolutional layers of the network seem perfectly fit to deal with simple patterns such as the moiré present in HRTEM images of CNTs.

These new image classification techniques have proven transformative in a number of fields and it can be expected that they will have a similar impact on electron microscopy. There is therefore, an emerging opportunity of bringing together recent advances in the fields of electron microscopy, atomistic simulation and deep learning, that we wish to seize for the determination of the chiral indices of CNTs. Up to now, very little research efforts went into the development of machine-learning-based analysis of electron microscopy images. General classification of scanning electron microscopy images of nanobjects in categories such as nanoparticles, nanowires, and nanopatterned surfaces has been achieved with artificial neural networks~\cite{Modarres2017}. The latter have also been trained for the segmentation of such images to automatically detect the position and orientation of CNTs~\cite{Trujillo2017}. Finally, local structures, such as defects or dopants, have been identified in TEM images by deep learning methods~\cite{Ziatdinov2017,Madsen2018,Li2018}.

In this contribution, we develop an image classification method based on CNNs for the determination of the structure of CNTs from experimental HRTEM images. Training of machine learning systems requires large data sets that we generate by means of computer simulation. This process is described in section~\ref{sec:training_data} and consists of two parts: The sampling of representative geometries of CNTs by molecular dynamics, and the generation of simulated HRTEM images based on these geometries by multi-slice simulations. Section~\ref{sec:classification_system} introduces our methodology for the chirality assignment of CNTs, consisting of an image preprocessing part and the CNNs for the image classification. The classification system is evaluated with the help of simulated images. In section~\ref{sec:application}, we show how the system applies to experimental HRTEM images. We conclude this paper, by discussing the implications and possibilities of further development of our approach in section~\ref{sec:conclusion}. The computer code of the analysis program is available from the corresponding author upon request.

\section{Generation of the training data}
\label{sec:training_data}

A large, reliable, and accurate training data set is crucial for machine learning. In principle, such data could be obtained from experimental HRTEM images, however, this approach is costly and the chirality of the CNTs, needs to be known. Therefore, we resort to the generation of the data set by numerical simulation. We use molecular dynamics to sample representative CNT geometries, which are in turn used for the generation of simulated HRTEM images.

\subsection{Molecular dynamics}

\begin{figure}[htb]
    \includegraphics[width=3.375in]{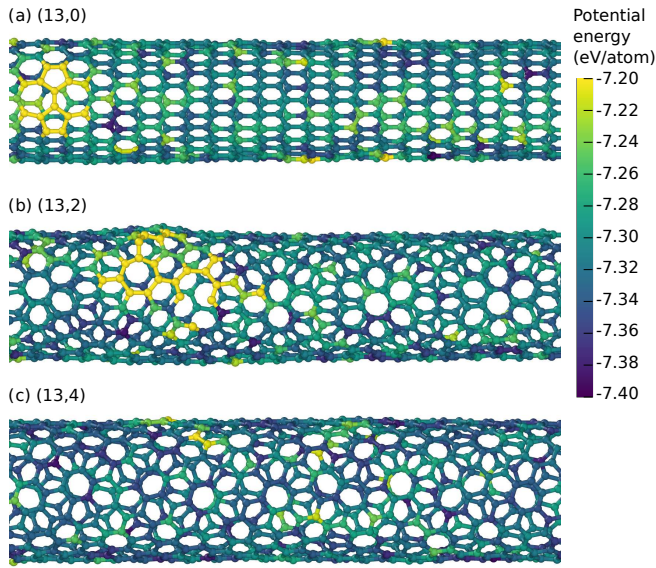}
    \caption{(color online) Some examples of final snapshots from the molecular dynamics simulations. Defects (higher potential energy, appear in yellow) have been introduced by high-temperature annealing. The visualizations are generated using the OVITO software~\cite{Stukowski2010}.}
    \label{fig:md_snapshots}
\end{figure}

Molecular Dynamics is employed to obtain representative configurations of CNTs in the diameter range of 0.48--2.30~nm using the LAMMPS simulation package~\cite{lammps,Plimpton1995}. This diameter range covers 261 chiralities, which are all considered in this study, and corresponds to CNTs produced by chemical vapor deposition techniques which serve as long-standing reference samples used in several works~\cite{Rao2018}. We use periodic boundary conditions in the direction along the tube axes. The length of the simulated CNTs is at least 6~nm containing up to 3316 atoms. The simulations rely on the Tersoff potential~\cite{Tersoff1989,Tersoff1990}, mostly because it has been successfully used to study the properties of carbonaceous structures~\cite{Meunier1998,Berber2000,Magnin2014} and allows for rapid simulations. The simulations are started at 6,500~K in order to statistically cause some defects in the walls of the CNTs. This way, a diverse and realistic reference data set is obtained which is important for the training of the deep learning system. Initially, the temperature is rapidly decreased to 300~K over 2,000 timesteps of $\Delta T=1.0$~fs. The simulation is then equilibrated at 300~K for another 8.000 $\Delta T$ before the final snapshot is extracted. Throughout the simulations, using a Nosé-Hoover barostat, the pressure is kept at zero. This simulation protocol is chosen to include some topological defects and realistic fluctuations of the atomic positions at ambient temperature. For each type of CNT, 1,000 independent simulation runs are carried out. Figure~\ref{fig:md_snapshots} shows a few examples of the final snapshots that are used as the basis for the generation of the simulated HRTEM images. The color code gives the potential energy according to the Tersoff potential, where a carbon atom in perfect graphene has an energy of -7.4~eV.

\begin{table}
\caption{Parameters for the generation of the HRTEM images. \label{tab:aberration}}
\begin{ruledtabular}
\begin{tabular}{c c}
Parameter                       & range                      \\ \hline
Geometrical parameters:                                      \\ \hline
Shift along tube axis           & 0  $\pm$  0.3 nm           \\
Rotation long tubes axis        & 0  $\pm$  180 $^\circ$     \\
Inplane tilt                    & 0  $\pm$  1.25 $^\circ$    \\
Out-of-plane tilt               & 0  $\pm$  1.25 $^\circ$    \\
Global coordinate scaling       & 1  $\pm$  0.025            \\
Electron microscope parameters:                              \\ \hline
High voltage & 80 kV \\
Defocus                         & 7  $\pm$  3.75 nm          \\
Twofold astigmatism A$_1$       & 5  $\pm$  5 nm             \\
Coma 3$\cdot$B$_2$              & 90 $\pm$  75 nm            \\
Threefold astigmatism A$_2$     & 30 $\pm$  25 nm            \\
Spherical aberration C$_3$      & 1  $\pm$  0.625 $\mu$m     \\
Star aberration, 4$\cdot$S$_3$  & 8  $\pm$  5 $\mu$m         \\
Fourfold astigmatism A$_3$      & 3  $\pm$  3 $\mu$m         \\
\end{tabular}
\end{ruledtabular}
\end{table}

In order to increase the diversity of the geometries, we apply some random geometric operations, also known as ``data augmentation'' in the context of deep learning. These include shifts in the direction of the tube axis, the rotation along the tube axis, the in-plane and out of plane tilt of the tube, and taking the mirror images of the atomic positions. Additionally, we apply a random scaling to the coordinates. This scaling is used to make the data set robust against small inherent systematic errors on interatomic distances in atomistic potentials.

\subsection{Simulation of HRTEM images}

Simulated HRTEM images have been calculated within dynamical theory with the Dr. Probe software package~\cite{Cowley1957,Barthel2018} using the multislice method~\cite{kirkland2010advanced}. We use this particular software, because of its integration with scripting languages such as bash and python, which makes the generation of a large amount of images straightforward. Object transmission functions for HRTEM simulations are computed on the basis of the geometries generated by molecular dynamics simulations. The geometries are partitioned into 6 equidistant slices along the electron beam. Electron scattering is computed for 80~keV electrons in a region of 3.6$\times$3.6~nm of the model potential at an inplane resolution of 75~px/nm. Thermal diffuse scattering is accounted for by Debye-Waller factors based on the absorptive potentials of Weickenmeier and Kohl~\cite{Weickenmeier1991}. In these simulations, only temporal coherence was taken into account, since for aberration corrected TEM imaging, the influence of spatial coherence can be neglected. This type of faster calculation is recommended for the simulation of images from Cs-corrected microscopes~\cite{Barthel2018}.

\begin{figure}[htb]
    \includegraphics[width=3.375in]{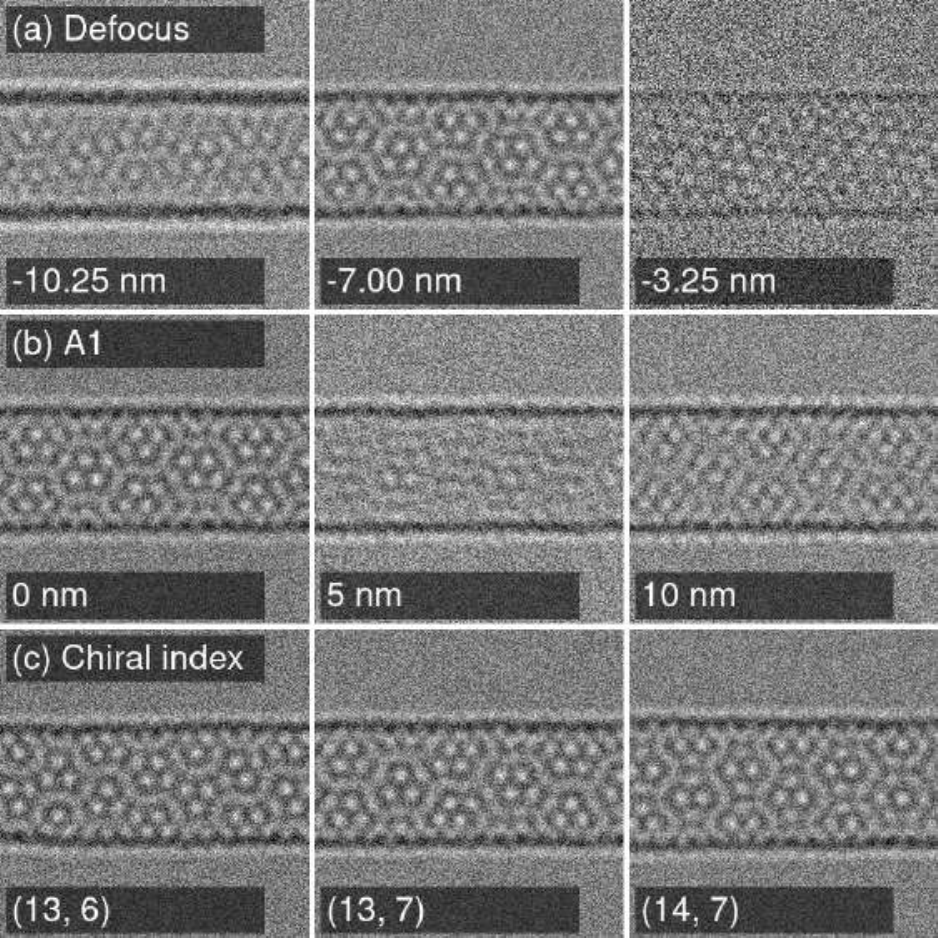}
    \caption{Effect of (a) defocus, (b) two-fold astigmatism (oriented at a 45$^\circ$ angle with respect to the CNT axis), and (c) chiral indices on the HRTEM image. Panels (a, b) show a (13,7) CNT. In all images, the defocus is set to -7~nm, the spherical aberration to 1~$\mu$m, and all other aberration coefficients to zero, except where indicated otherwise.}
    \label{fig:aberration}
\end{figure}

In order to further increase the diversity of the training data set, we apply random aberration coefficients for the generation of the images. The full list and the range of variation of these parameters are given in Tab.~\ref{tab:aberration}. The mean values of the aberration coefficients correspond to the objective lens of the JEOL JEM-ARM200F spherical-aberration-corrected electron microscope equipped with a cold field emission gun operated at 80 kV~\cite{Ricolleau2013} used for the acquisition of the experimental images (see section~\ref{sec:application}). The in-plane orientation of the respective aberration coefficients is also randomly selected in the interval of 0 to $2\pi$. Final images are obtained by normalizing, equalizing, adding a white noise to the result and reducing the resolution to 26.7~px/nm. This resolution is close to the lowest possible one that still preserves most of the moiré features present in the CNT HRTEM images. Some examples can be found in Fig.~\ref{fig:aberration}. Note, that while the aberration coefficients have strong effects on the TEM contrast [(panels (a, b)], the differences in the moiré patterns of CNTs of different chiral indices can be virtually invisible to the naked eye [panel (c)]. Both of these observations make the classification task of the images in terms of chiral indices complex. Finally, from each of the generated images, one random 64$\times$64~px subimage is randomly selected along the tube's axis for the training of the CNNs.

\section{Architecture and training of the classification system}
\label{sec:classification_system}

In the following, we describe our automatic classification system that determines the chiral index of a CNT based on a HRTEM image. We use an analytical procedure for the alignment of the tube followed by the classification by a CNN trained on the data set generated by means of numerical simulation.

\subsection{Architecture and training of the CNNs}

\begin{figure}[htb]
    \includegraphics[width=3.375in]{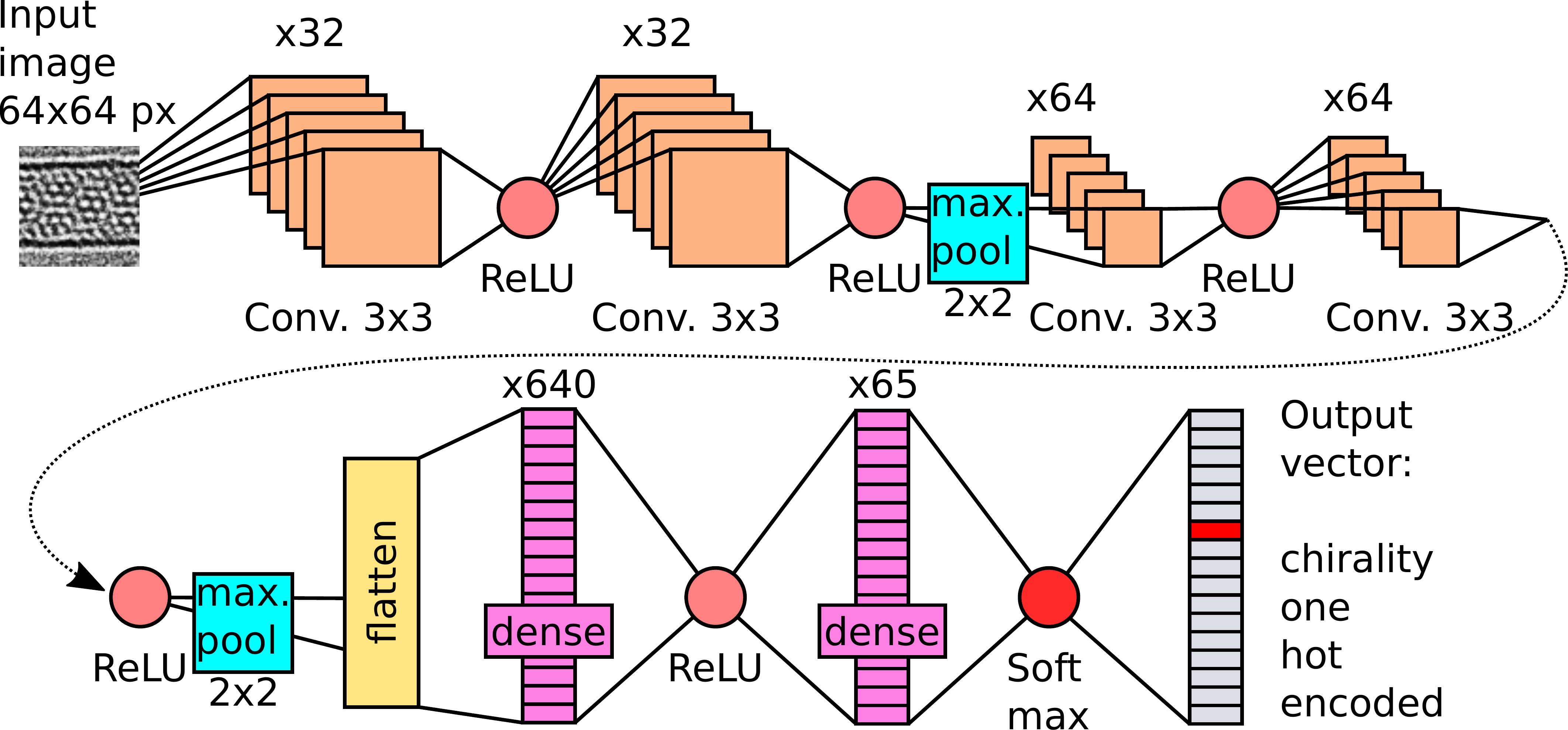}
    \caption{(Color online) Schematic detailing the architecture of the CNNs.}
    \label{fig:cnn_architecture}
\end{figure}

In order to simplify the classification task, a first CNN is used to group the CNTs by diameter into one of nine fully overlapping sets. This way, the maximum number of chiralities an individual CNN needs to classify is reduced to 65 (down from 261). However, this requires the training of one additional CNN that determines the diameter class. According to the result of the first classification, the image is then analyzed by a second CNN that determines the chiral index within the respective diameter class.

The architecture of the CNNs used for this work is based on LeNet-5~\cite{Lecun1998}. Using the keras software~\cite{Chollet2015}, we define a network comprising four convolutional layers followed by two fully connected layers, a schematic representation is shown in Fig.~\ref{fig:cnn_architecture}. Each of the layers uses rectified linear units as activation functions, except for the final classification where a normalized exponential function (softmax) is used in order to interpret the results as probabilities for particular chiralities. Additionally, after the second and fourth convolutional layer, 2$\times$2 max-pooling layers are included for downsampling and increased tolerance for translation in the input images. The first two convolutional layers use 32, and the second two use 64 3$\times$3 filters (of stride 1). The first fully connected layers contains 640 output neurons, while the number of output neurons of the second fully connected layer is determined by the number of output classes of the respective CNN. The total number of adjustable parameters per CNN is thus $\approx 10^6$.

\begin{figure}[htb]
    \includegraphics[width=3.375in]{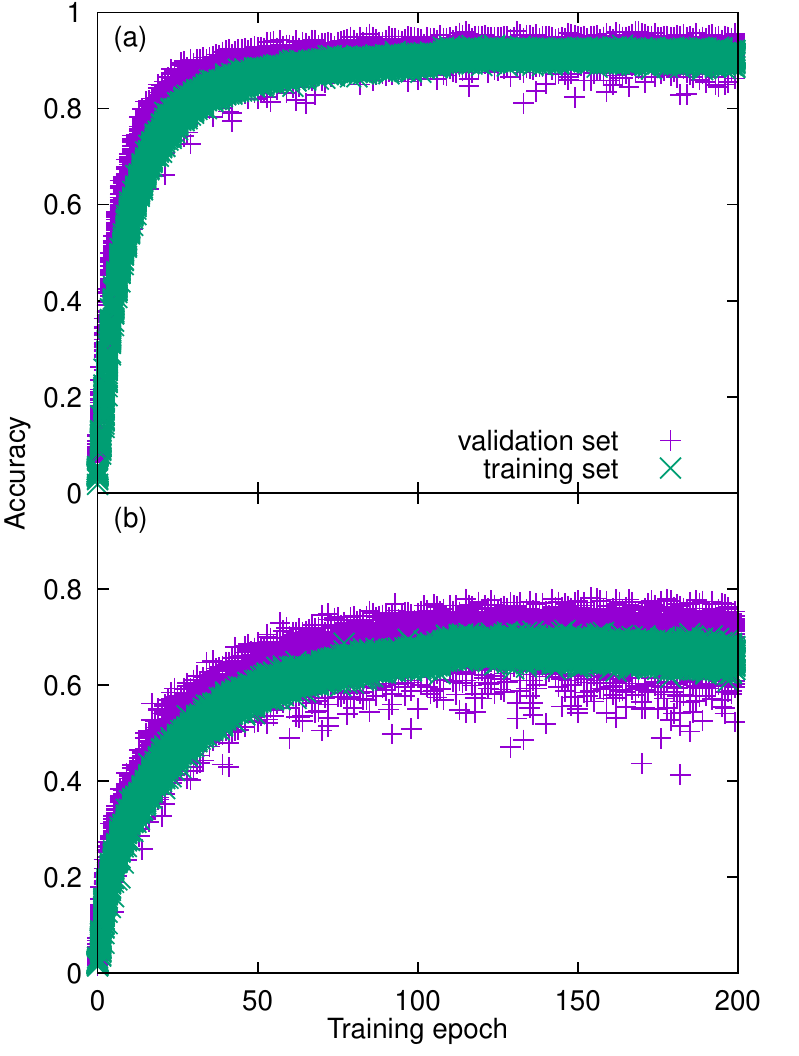}
    \caption{(Color online) Accuracy during training of CNNs concerned with CNTs in the diameter range of (a) 0.48--1.19~nm and (b) 1.98--2.28~nm, each containing 65 chiralities.}
    \label{fig:training_accuracy}
\end{figure}

For the training of the CNNs, the simulated images are split in a 4:1 ratio into a training and validation set. In order to enhance training, a random shift in image brightness of max. $\pm 15\%$ is applied to the images in the training set. During training, a dropout layer (25\%) was added before the final fully connected layer~\cite{Srivastava2014}. This randomly sets one quarter of the input units to zero at each update during optimization which helps to prevent overfitting, i.e., the phenomenon of the network performing well only on the training data set, but not on the validation set. Training the network, i.e. adjusting the $\approx 10^6$ parameters, is achieved with the root mean square propagation (RMSprop) algorithm. The parameters are optimized with respect to categorical cross-entropy as the objective function.

Figure~\ref{fig:training_accuracy} shows the classification accuracy during the training process. Due to the dropout layer, the accuracy during training is artificially reduced and thus lower than the classification accuracy of the validation set. It appears that the training yields best results for the low-diameter CNTs. This is presumably due to the increasing similarity of the moirés from the larger tubes. For each of the nine CNT diameter classes, 40 instances of the CNN are trained. Combination of the classification results of these 40 instances, which are admittedly rather correlated, does allow for a moderate increase in overall classification accuracy.

\subsection{Image preprocessing}

\begin{figure}[htb]
    \includegraphics[width=3.375in]{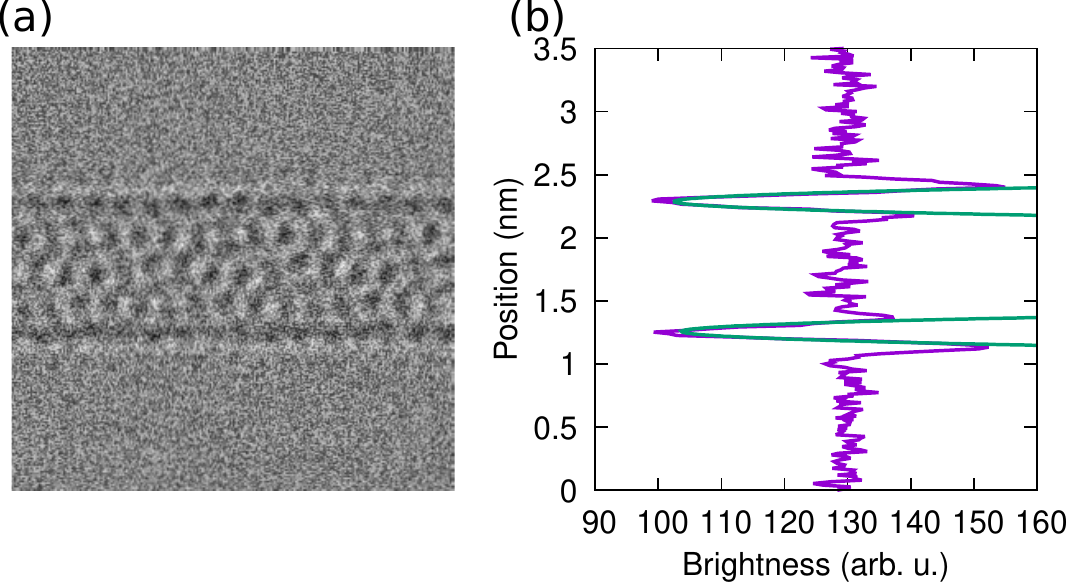}
    \caption{(Color online) (a) Simulated HRTEM image of a (10,5) CNT, (b) Projection of brightness of the image perpendicular to the CNT axis for the analytical alignment and diameter determination via parabolic fits at the location of the brightness valleys of the Fresnel fringes.}
    \label{fig:preprocessing}
\end{figure}

Before analyzing HRTEM images by the CNNs, they are processed first using the following procedure. In a first step, the images are oriented in such a way that the axis of the tubes are horizontally aligned. This is achieved by projecting the grayscale values of the images as a function of image orientation. The Fresnel fringes from the sides of the CNTs correspond to minima in the projection curves. The angle of the projection where the minima are deepest and narrowest aligns the tube (see Fig.~\ref{fig:preprocessing}).

In a second step, the images are downsampled to the image resolution used during the training of the CNN, i.e., 26.7~px/nm. From the rescaled images 40, 64$\times$64~px subimages are cropped, normalized and equalized. These subimages are then processed first by the CNN that assigns them a particular class of diameter. After that, they are analyzed by the 40 instances of the CNN corresponding to the CNT diameter class determined by the first CNN. Finally, from the resulting total of 40$\times$40 chirality estimates an overall chirality prediction is aggregated.

\subsection{Evaluation of the classification system}
\label{sec:evaluation}

\begin{figure}[htb]
    \includegraphics[width=3.375in]{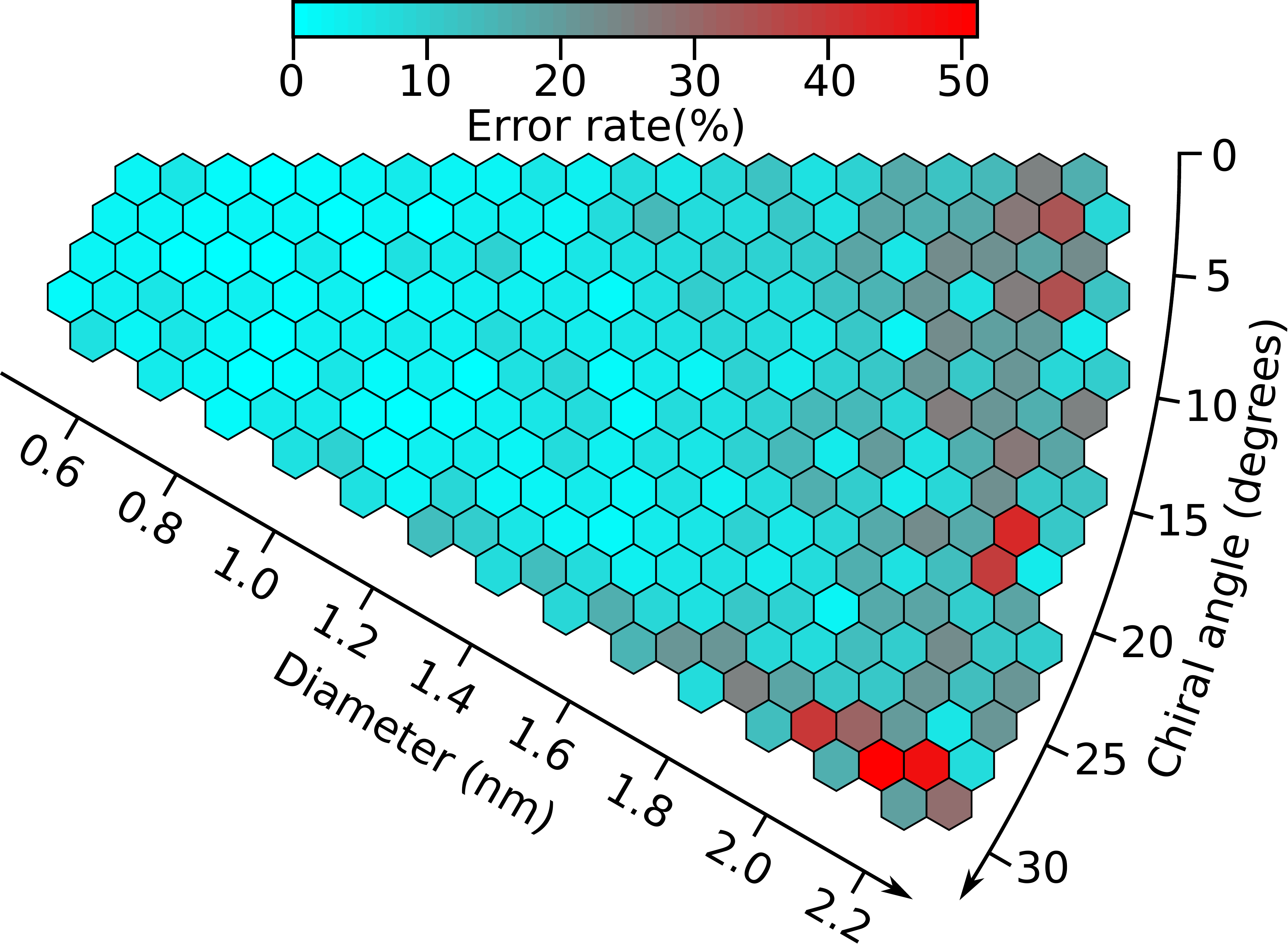}
    \caption{(Color online) Rate of misclassification by the combined assignment system as a function of CNT chirality.}
    \label{fig:error_rates}
\end{figure}

The performance of the classification system is first evaluated using simulated HRTEM images. For this purpose, we take the entire system including the preprocessing stage and use 100 images of each of the considered chiralities containing the same variability as the training data. Figure~\ref{fig:error_rates}, shows the rate of misclassification as a function of chiral angles and diameters. It turns out that the first CNN that determines the diameter class correctly classifies the image in over 99\% of the cases, while the overall system reaches an accuracy of 90.5\%, averaged over all chiralities. This means that most of the errors occur with the second CNN that assigns the final chiral indices. The overall system performs best for low diameter tubes. This tendency can be understood easily, because for low diameter tubes, there are less possible tubes in a given diameter range, and the possible chiral angles differ more between neighboring chiral indices.

For CNTs of larger diameters, misclassifications are more likely in the case of the near armchair tubes. The unit cells of CNTs tend to grow with diameter, and therefore in the individual small 64$\times$64 px images that are the inputs for the CNNs, only a relatively small fraction of the moiré pattern is present which makes the classification task more difficult.

\begin{figure}[htb]
    \includegraphics[width=3.375in]{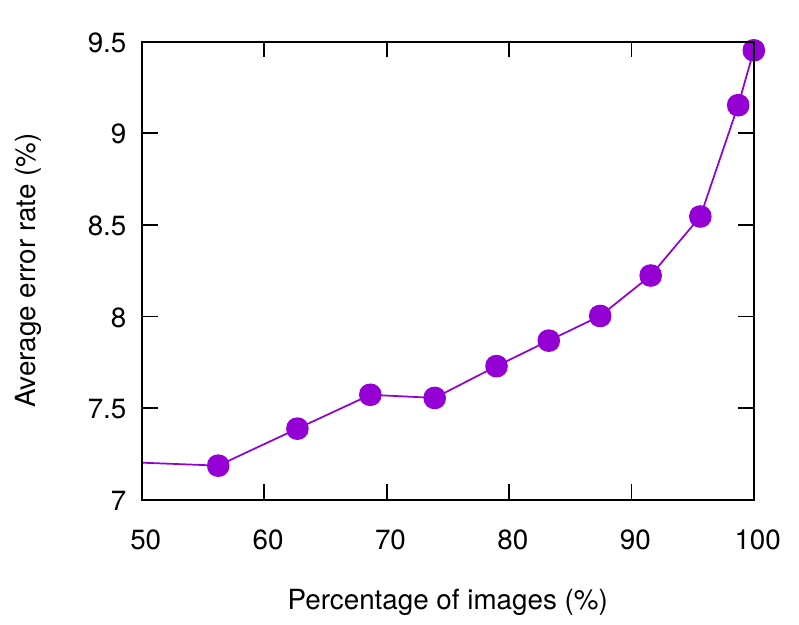}
    \caption{(Color online) Classification accuracy as a function of image quality percentile. Image quality is quantified as the contrast between the bright and dark fringes at the sides of the CNT.}
    \label{fig:quality2}
\end{figure}

Furthermore, it seems reasonable to assume that the image quality has an effect on the classification accuracy. In order to discuss this effect, the image quality of the 27100 images from the test set has been quantified. For this purpose, the ratio of the intensity of the bright and dark fringes at the sides of the tubes has been evaluated (see Fig.~\ref{fig:preprocessing}). Ordering the images by this measure of images quality, it turns out that the overall error rate can be reduced by taking into account only the images with the highest contrast between the dark and bright fringes. Figure~\ref{fig:quality2} shows the classification accuracy of the overall system as a function of the percentile of the highest quality images. This emphasizes that the classification methodology relies on high quality images.

\begin{figure}[htb]
    \includegraphics[width=3.375in]{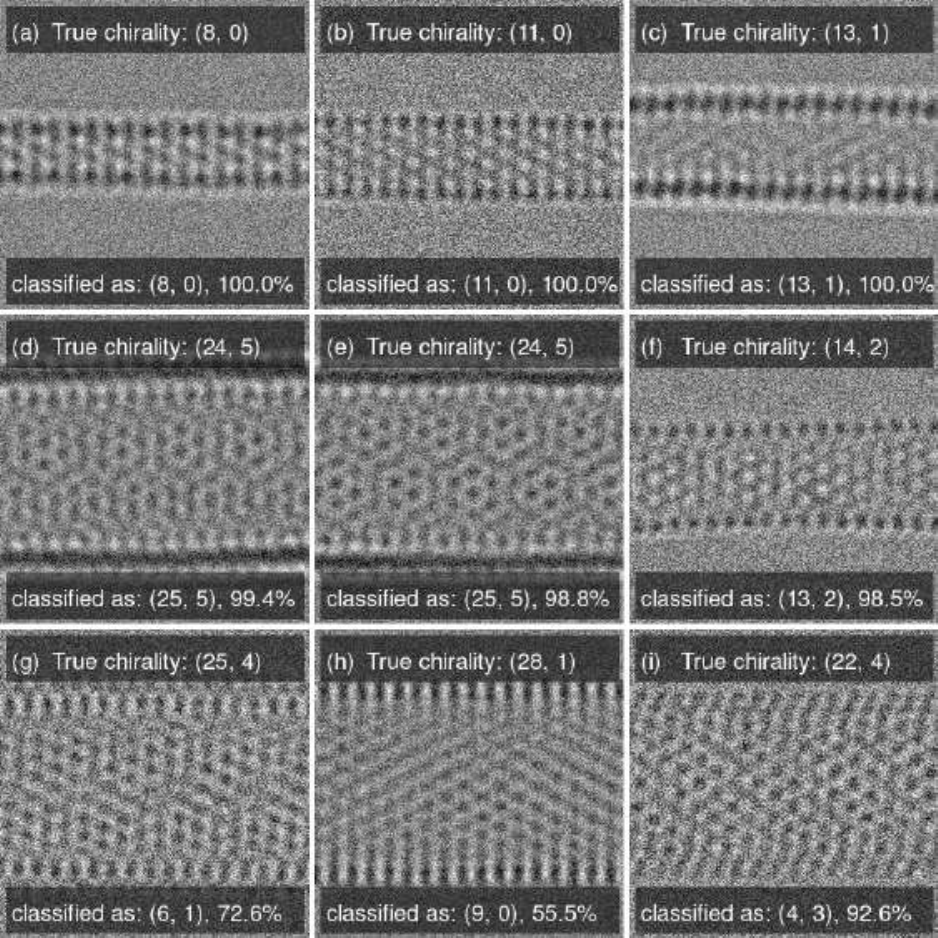}
    \caption{Examples of assignments of chiral indices of CNTs in the case of simulated HRTEM images. (a-c) Correctly classified images with high assignment confidence. (d-f) Wrongly classified images with a high confidence. These images mostly correspond to CNTs with an above-average defect density. (g-i) Wrongly classified images where the predicted chiral indices are far from the correct one.}
    \label{fig:error_cases}
\end{figure}

Figure~\ref{fig:error_cases}(a-c), gives examples of images that resulted in correct classifications with high probability. Unsurprisingly, this occurs most often with images containing clear moiré patterns of CNTs with small diameters. In such cases, the result is very clear and allows unambiguous identification of the chirality of the considered tube. It is also instructive to examine individual cases where the classification system fails. In some cases, the system indicates high probabilities ($>$95\%) for a certain wrong chirality [see Fig.~\ref{fig:error_cases}(d-f)]. Most of these misclassifications occur for structures with clearly visible defects and the predicted chiral index $n$ differs by only one unit from the true index. This type of error tends to occur more often with CNTs of larger diameter. Figure~\ref{fig:error_cases}(g-h) shows examples of misclassifications where the differences between the predicted and true chiral indices are highest. These error cases occur when the first CNN that determines the diameter range of the CNT fails. In this case, the final analysis is carried out by a CNN that does not include the correct chirality in its classification range. Interestingly, most often, the wrong CNN will still select a chirality with a similar chiral angle as the one of the correct CNT. In these cases, the defocus mostly has rather extreme values, which makes the fringes at the sides of the CNTs as well as the moiré pattern less clear. These are conditions that makes the chirality assignment difficult also for human experts. It turns out, however, that this error mode never occurred when analyzing experimental HRTEM images [see Fig.~\ref{fig:application}(b)]. Several alternative approaches to programmatic chirality assignment are discussed in section~S1.

The range of variations of geometrical aberrations used for the training of the CNNs may seem very narrow. Indeed, aberrations may change over time~\cite{Barthel2013} and oscillation of the sample under beam irradiation affects the focal distance~\cite{Alloyeau2012}. We assume thus that some experimental images analyzed in this work have been acquired with optical parameters slightly out of the ranges defined in Tab~\ref{tab:aberration}. It turns out that our classification system still works well rather far outside these ranges, as discussed in section~S5. We presume that this is due to the fact that visually the effect of different aberration coefficients can be rather similar. Therefore, due to the randomization of aberration coefficients in our training set, extreme images, similar to the ones of section~S5, are also included in our training data base.

\section{Application to experimental HRTEM images}
\label{sec:application}

The performance of the classification system was then evaluated on experimental HRTEM images. CNTs were observed using an aberration corrected Jeol ARM 200F microscope operating at 80 kV. Images from three different samples were used: CVD-grown CNTs prepared based on the method described in Ref.~\onlinecite{Castan2017} using an iron catalyst, and two chirality-enriched samples obtained from a HiPCo sample (NoPo Nanotechnologies, Inc.) by the aqueous two-phase extraction method~\cite{subbaiyan2014role, VanBezouw2018}. In the following, we will first present the established manual approach to determine the structure of a CNT from the experimental images (adapted from Ref.~\onlinecite{Ghedjatti2017}). Then we compare with the chiral index assignment by our automatic procedure.

\begin{figure*}[htb]
    \includegraphics[width=6.4in]{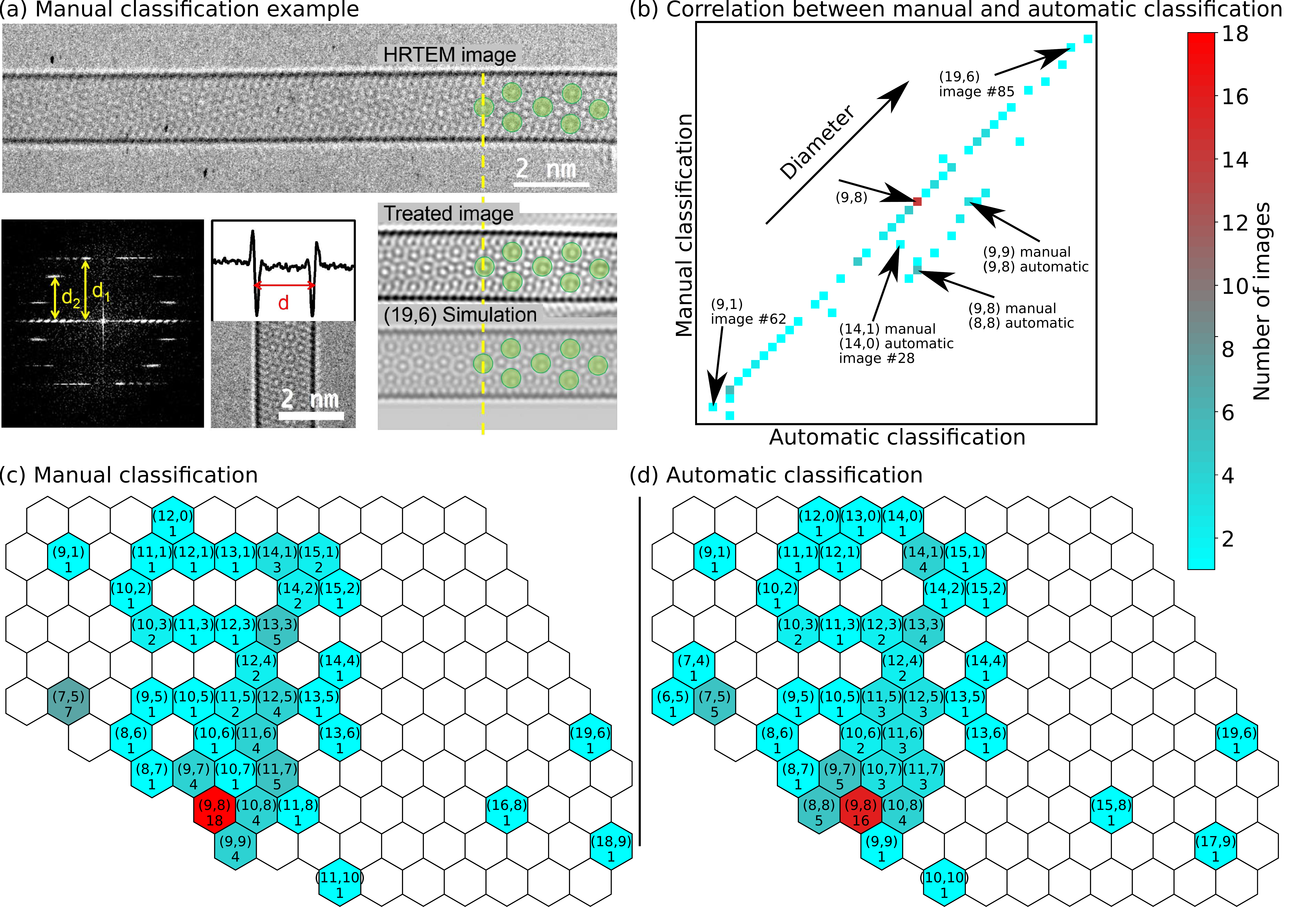}
    \caption{(Color online) Comparison of the results of the conventional (manual) and automatic (deep learning) based chirality determination methods. (a) Example (CNT 85 in Tab.~S1) of a manual classification, agreeing with automatic classification: experimental HRTEM image (top), extracted FT of the image for chiral angle measurement (13.5 $\pm$ 0.8 $^{\circ}$) and perpendicular intensity profile for diameter measurement (1.82 $\pm$ 0.05 nm) (bottom left). These measurements lead to one possible chirality which is (19,6). The experimental image, treated by applying a mask on the FT showing only SWCNT layerlines, is compared to a simulated (19,6) image (bottom right), confirming the classification. (b) Correlation matrix between the classification results of the two methods. (c,d) Results of the classifications of 91 experimental images, using the manual and automatic methods, respectively. The number of images for each chirality is color-coded and indicated below the chiral indices.}
    \label{fig:application}
\end{figure*}

First, the diameter of the nanotube was measured by extracting a contrast intensity profile perpendicular to the tube axis [see Fig.~\ref{fig:application}(a)]. According to image simulations, the diameter at Scherzer defocus is equal to the distance between the two inflection points between the dark and bright fringes on the sides of the CNT, meaning the midpoint between the minimum of the dark fringe, and the maximum of the subsequent bright fringe. A Fourier Transform (FT) of the HRTEM image is then used to determine the chiral angle ($\theta$) of the nanotube as discussed in section~S1 [see also Fig.~\ref{fig:application}(a)]. Considering the measured diameter and chiral angle values (including error bars), we can assign possible $(n,m)$ pairs for the CNT. Comparing the FT with a calculated diffraction pattern for these chiralities can help exclude some possibilities. To discriminate between the possible $(n,m)$ pairs, the HRTEM image is compared to image simulations of all the possible chiralities. The comparison of the moiré patterns leads to an unambiguous $(n,m)$ assignment.

For the automatic determination, a rectangular region of interest is selected from the experimental image which is then analyzed by the classification system. The results for the two methods were compared for a total of 91 CNT images. Only images of sufficient quality were used for this statistical analysis: images where nanotubes were not fully suspended, too contaminated or damaged, or where the focus and alignment conditions of the microscope did not allow the extraction of a readable FT, were not included.

The analytical process of manually determining the structure of a tube is of the order of 15 to 30 minutes. In comparison, depending on computing hardware, the analysis using the deep learning method is much faster and can be carried out without supervision. The resulting chiralities using both, the conventional and deep learning based chirality determination methods, is shown in Fig.~\ref{fig:application}(c,d), respectively, and the correlation of the classification methods is represented in Fig.~\ref{fig:application}(b). Cases where the two methods lead to identical results correspond to elements on the diagonal. Elements below and above the diagonal indicate images that have been classified as a chirality of a smaller and larger diameter with the deep learning method than with the manual method, respectively. Overall, the two methods lead to the same result in 71 $\%$ of cases. An example of a manual classification in a case where the two methods agree, is shown in Fig.~\ref{fig:application}(a). The individual results are given in section~S2, some more successfully classified images in section~S3, and an example where the automatic procedure fails in section~S4. When there is a disagreement between the two methods, the chirality that was found conventionally is the second or third most probable chirality according to the deep learning tool in three out of four cases. In most of the error cases, the mismatch in chiralities is always minimal: if the manual technique gives an $(n,m)$ chirality, then the automatic technique will likely give (n $\pm$ 1,m), or (n,m $\pm$ 1). When removing the nanotubes for which the automatic method is indecisive (probability for the first chirality below 80 $\%$) from the data set, the agreement percentage goes up to 100 $\%$.

\section{Conclusion}
\label{sec:conclusion}

We have developed a robust and efficient tool to obtain in a very simple way the CNT chirality from raw HRTEM images. Based on a classification method using CNNs, the determination of the chirality of CNTs is fast enough to provide meaningful statistics on experimentally produced samples. An important and original point of our approach is the establishment of a large database obtained from atomic-scale simulations to train the CNNs. The consideration of defects makes the images of our training data base comparable to experimental data.

Such an easy-to-use tool will be of great interest to the CNT community in different research fields. A typical case is the lack of control of CNT chirality during synthesis, which is a major obstacle to the widespread use of CNTs in technological applications~\cite{Yang2014,Amara2017,Magnin2018,Rao2018,Forel2019}. This critically depends on the reliable analysis of the chirality distribution of CNT samples. Another approach to achieve chiral selectivity within CNT samples is to work on the subsequent processing and sorting of the raw material~\cite{Arnold2006,hersam2008progress,fagan2019aqueous}. Here again, the validation of the sorting methods requires a precise analysis in terms of chirality, which can be performed using our tool.

There is no doubt that our approach will be useful for other fields of research related to nanotubes but also in nanoscience more generally. The approach implemented can clearly be adapted to other nano-objects such as the structure of nanoparticles (pure, bimetallic, twins)~\cite{Prunier2015}, characterization of defects in 2D materials~\cite{Mouhoub2020} or identification of stacking in Van der Waals heterostuctures~\cite{Banhart2011, Rasool2015, Geim2013}. In addition, the implementation of the new generation of so-called sub-angstrom low-voltage electron microscopes makes it possible to study radiation-sensitive materials at ultra-high resolution~\cite{Cao2020}. Our tool will be very beneficial for the analysis of the resulting large amount of very accurate data~\cite{Shree2019}. Our approach demonstrates the great potential of deep learning methods for the analysis of HRTEM images, and we hope that it will stimulate further developments in this direction, such as single image super resolution~\cite{Lim2017}, in the very near future. In contrast to super resolution in other domains, the relevant physics of nanoscale systems is essentially known and may be transferred to the deep learning model during training.

\begin{acknowledgments}
The authors thank the METSA research network for giving access to the Cs-corrected TEM of the MPQ-Paris Diderot laboratory. The authors acknowledge M. Kociak (DiffractX) and Y. Le Bouar for use of their software, and S. Cambré, W. Wenseleers, and J. Defillet for providing part of the samples used for experimental images. This work benefited from the support of the project ANR GIANT (project 18-CE09-0014) of the French National Research Agency (ANR).
\end{acknowledgments}

\bibliography{bibliography}
\end{document}


\title{Supplemental material\\A deep learning approach for determining the chiral indices of carbon nanotubes from high-resolution transmission electron microscopy images}

\author{G. D. Förster}
\affiliation{Laboratoire d'Etude des Microstructures, ONERA-CNRS, UMR104, Université Paris-Saclay, BP 72, 92322 Châtillon Cedex, France}
\affiliation{Aix Marseille Univ, CNRS, CINaM, Marseille, France}
\author{A. Castan}
\affiliation{Laboratoire d'Etude des Microstructures, ONERA-CNRS, UMR104, Université Paris-Saclay, BP 72, 92322 Châtillon Cedex, France}
\affiliation{Department of Physics and Astronomy, University of Pennsylvania, Philadelphia, PA 19104, USA}
\author{A. Loiseau}
\affiliation{Laboratoire d'Etude des Microstructures, ONERA-CNRS, UMR104, Université Paris-Saclay, BP 72, 92322 Châtillon Cedex, France}
\author{J. Nelayah}
\affiliation{Université de Paris, Laboratoire Matériaux et Phénomènes Quantiques, CNRS, F-75013, Paris, France}
\author{D. Alloyeau}
\affiliation{Université de Paris, Laboratoire Matériaux et Phénomènes Quantiques, CNRS, F-75013, Paris, France}
\author{F. Fossard}
\affiliation{Laboratoire d'Etude des Microstructures, ONERA-CNRS, UMR104, Université Paris-Saclay, BP 72, 92322 Châtillon Cedex, France}
\author{C. Bichara}
\affiliation{Aix Marseille Univ, CNRS, CINaM, Marseille, France}
\author{H. Amara}
\affiliation{Laboratoire d'Etude des Microstructures, ONERA-CNRS, UMR104, Université Paris-Saclay, BP 72, 92322 Châtillon Cedex, France}
\affiliation{Université de Paris, Laboratoire Matériaux et Phénomènes Quantiques, CNRS, F-75013, Paris, France}

\email{daniel.foerster2@gmail.com}

\date{\today}

\maketitle

\beginsupplement

\section{Alternative approaches of chirality assignment}
\label{sec:alternatives}

As an alternative to the deep learning approach presented in the main text, several other methods of automatized chirality determination of CNTs from HRTEM images have been explored as part of the present effort. They have however been less successful. For the sake of completeness, they are briefly introduced in the following paragraphs.

\begin{figure}[htb]
    \includegraphics[width=3.375in]{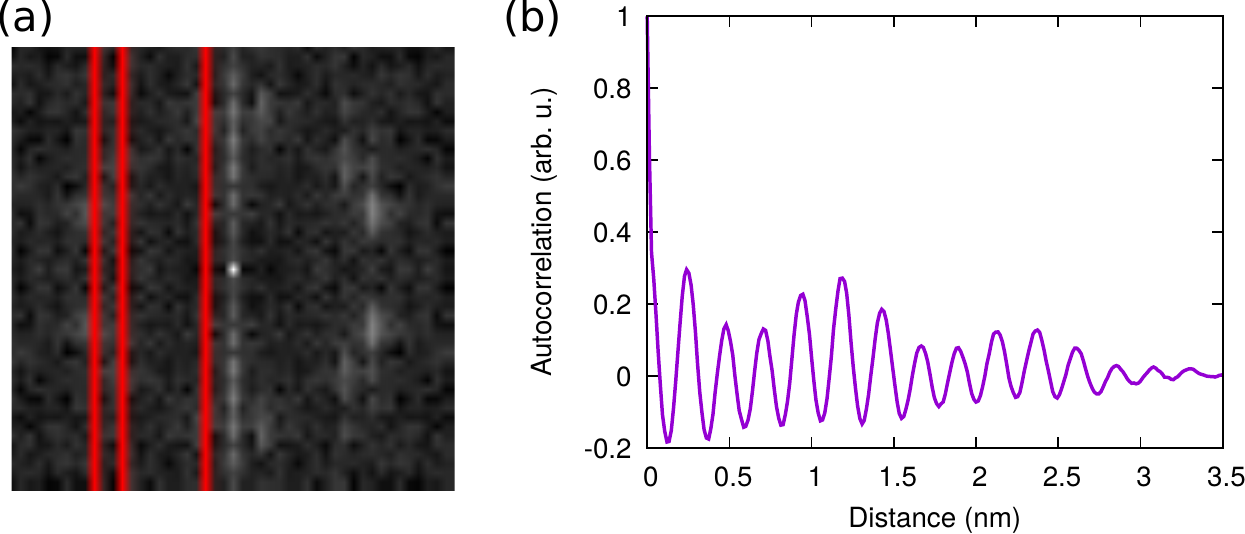}
    \caption{(Color online) (a) Fourier transformation (FT) of a simulated image of a (10,5) CNT (FT from Fig.~5(a) main text) with layer lines highlighted in red, and (b) autocorrelation of brightness along the CNT axis.}
    \label{fig:alternatives}
\end{figure}

\paragraph{Analytical estimation of CNT diameter and chiral angle}

It is possible to recover the chirality of a CNT with analytical procedures for the determination of CNT diameter and chiral angle from the HRTEM images. In HRTEM images, the CNT diameter is usually taken as the distance between the inflection points between the dark and the bright fringes~\cite{Fleurier2009}. However, the maxima from the bright fringes are often not intense enough to allow for a straightforward automatic detection [see Fig.~5, main text]. It is also possible to use only the dark fringes, and add a constant offset to their separation in order to estimate the CNT diameter. We estimated this offset as 0.06~nm between the separation of the dark fringes and the CNT diameter from the comparison of the diameter obtained from a number of simulated HRTEM images with the theoretical value. Using this procedure, the diameter of the (10,5) CNT of Fig.~5 (main text) is estimated at 1.09~nm, while the theoretical value amounts to 1.05~nm. The diameter estimate is used here only to select the appropriate CNN that has been trained for the diameter range the particular CNT falls into. This purpose does not require a very precise diameter estimation.

The chiral angle can be extracted from the Fourier-transformed (FT) HRTEM image of the CNT [Fig.~\ref{fig:alternatives}(a)]. This requires determining the so-called layer lines~\cite{Zuo2003}, which correspond to the three lines of maximal brightness in the FT image (highlighted in red on the figure). As with the CNT diameter, these lines can be detected by parabolic curve fits to the projected intensity profiles. The chiral angle $\Theta$ is then calculated as a function of their distances $d_3<d_2<d_1$ from the equatorial lines:
$$\Theta=\arctan\left(\frac{d_2-d_3}{\sqrt{3}d_1}\right)=\arctan\left(\frac{2d_2-d_1}{\sqrt{3}d_1}\right)$$
For the example in Fig.~\ref{fig:alternatives}(a), the layer lines are $d_3=3$~px, $d_2=12$~px, and $d_1=15$~px away from the equatorial line, which leads to an estimation for the chiral angle of 19$^\circ$, which is compatible with the (10-5) chirality of $\Theta=19.1^\circ$.

However, in practice, both the diameter as well as the chiral angle determinations are not precise enough for an automatic assignment of the chirality. In addition, chiral angles can be determined with reasonable accuracy only if larger portions (compared to what is necessary for the analysis with the CNNs) of the CNT are visible or when the electron diffraction pattern is available.

\paragraph{Method based on CNT periodicity} Another method to assign the chirality is the determination of the periodicity along the CNT axis. In the case of some chiralities, this periodicity is not unique to a single chirality [see Fig.~\ref{fig:alternatives}(b)]. This means that this approach needs to be complemented by information either on the diameter or on the chiral angle in order to yield a unique chirality. Ideally, the autocorelation function shows a maximum after one period. However, several secondary maxima with similar intensity may also be present, which makes the (automatic) selection of the correct maximum difficult. This procedure requires that at least one entire period is visible in the image which may not always be the case. It should be noted that the largest unit cell in the present selection of CNTs in the diameter range of 0.48--2.30~nm is 12.4~nm. In the case of the example of Fig.~\ref{fig:alternatives}, the maximum is found at 1.2~nm, which should be compared to the theoretical value of 1.1~nm for (10-5) CNTs.

\paragraph{FT-based methods} Alternatively, the chirality could be determined by using only the FT images. The transformed images contain all the necessary information for the chirality determination: The chiral angle is encoded in the layer lines in these images, but also the diameter information is present on the equatorial line. Based on this fact, two methods were explored in this work: Firstly, FT images of the previously introduced simulated HRTEM images were computed for all the considered nanotube chiralities. Aggregates of these images can then be compared to the FT of the images with the unknown chirality. This task can be automatized by computing the correlation of the reference images with the unknown image (the pixel by pixel dot product of the grayscale value of the two images). With this approach, the chirality of the reference FT image of the highest correlation corresponds to the chirality prediction. Tests on our training set showed that error rates are much higher than what was achieved with the CNN-based method. Secondly, we attempted to train the CNN on the FT images. This was largely without success and the predictions were not significantly better than random guesses. We speculate that this is due to the fact that, in the case of the FT images, the information relevant for the chirality prediction is concentrated in a small number of pixels while much of the image is noise.

\section{Comparison of classification results from the conventional and the deep learning method}
\label{sec:comparison}

Table~\ref{tab:comparison} shows the results of the chirality determination of 91 individual CNTs from HRTEM images using the conventional and deep learning methods. The conventional method uses the measurements of the diameter using the Fresnel fringes and the chiral angle from the layer lines in the FT images. The results of these measurements are shown in the table. Assuming uncertainties of 0.5~\AA\ on the diameter and of 0.5 - 3.0$^\circ$ on the chiral angle, between one and four candidate chiralities need to be inspected in more detail. Comparing the moiré pattern and the FT pattern with simulated reference images leads to the final estimation of the chiral indices. The deep learning result corresponds to the aggregated probability of 1600 chirality estimates (40 CNN instances times 40 subimages). The three most likely chirality predictions together with their probabilities are given in Tab.~\ref{tab:comparison} for each of the images. It should be noted, however, that the probabilities given by the deep learning tool can vary slightly depending on the selected section for analysis. Close inspection [compare also with Fig.~8(b), main text] shows that the conventional method tends to yield larger diameters than the deep learning method. Contributing reasons may include a systematic error in the numerical simulations due to the interatomic model that are used to populate the training data base, or a systematic error on the diameter estimation from the Fresnel fringes used with the conventional method. Furthermore, it turns out that disagreements between the two methods are more likely in the case of chiral angles close to 0$^\circ$ or 30$^\circ$. However, there seems to be no systematic over- or underestimation of chiral angles. On the one hand, Fig.~6 (main text) indicates that the CNNs have more difficulties with correct chirality predictions in these cases. On the other hand, the differences in FT patterns are visually most clear at chiral angles close to zero. These two observations show that a manual inspection of these cases may help to improve on the automatic analysis. In case of large chiral angles, close to 30$^\circ$, the definitive chirality determination seems to be most difficult with either of the methods.

\begin{table*}[ht]
\centering
\caption{Comparison of classification results from the conventional and deep learning methods on experimental HRTEM images acquired from 91 SWCNTs. For the conventional method (``Manual''), the measured diameters and chiral angles, and resulting chirality are given. For the deep learning method (``Automatic''), the three most probable chiralities, along with their corresponding probabilities are given.}
\label{tab:comparison}
\begin{tabular}{c|c|c|c|c|c|c|c|c|c|c|c|c|c|c|c|c|c|c|c|c}

 & \multicolumn{3}{c|}{Manual} & \multicolumn{6}{c|}{Automatic} &  &  & \multicolumn{3}{c|}{Manual} & \multicolumn{6}{c}{Automatic} \\ \hline
CNT & d (nm) & $\Theta$ ($^\circ$) & (n,m) & (n,m)1 & P1 & (n,m)2 & P2 & (n,m)3 & P3 &  & CNT & d (nm) & $\Theta$ ($^\circ$) & (n,m) & (n,m)1 & P1 & (n,m)2 & P2 & (n,m)3 & P3 \\ \hline
1 & 1.20 & 16.1 & 12,5 & 11,5 & 78.7 & 12,5 & 15.3 & 12,4 & 3.6 &  & 47 & 1.14 & 16.7 & 11,5 & 11,5 & 59.8 & 12,5 & 18.9 & 11,6 & 12.4 \\ \hline
2 & 1.18 & 27.8 & 9,8 & 9,8 & 74.1 & 8,8 & 14.5 & 9,7 & 9.1 &  & 48 & 1.24 & 30.0 & 9,9 & 9,8 & 50.6 & 9,7 & 34.1 & 8,8 & 7.4 \\ \hline
3 & 1.21 & 16.1 & 12,5 & 12,5 & 78.2 & 12,4 & 11.3 & 11,5 & 7.9 &  & 49 & 1.27 & 22.9 & 11,7 & 11,7 & 77.9 & 11,6 & 18 & 10,7 & 1.7 \\ \hline
4 & 1.05 & 11.5 & 11,3 & 11,3 & 98.8 & 10,4 & 0.7 & 10,3 & 0.4 &  & 50 & 1.22 & 21.4 & 11,6 & 11,6 & 78.2 & 12,6 & 8.6 & 10,6 & 5.4 \\ \hline
5 & 1.19 & 3.2 & 14,1 & 14,1 & 72.4 & 14,2 & 21.2 & 13,2 & 5.3 &  & 51 & 1.19 & 10.9 & 13,3 & 13,3 & 94.5 & 13,2 & 3.7 & 12,3 & 0.8 \\ \hline
6 & 1.20 & 26.6 & 9,8 & 9,8 & 49.2 & 9,7 & 18.6 & 8,8 & 15.5 &  & 52 & 1.20 & 27.0 & 9,8 & 9,8 & 56.4 & 9,7 & 20.1 & 10,7 & 13.4 \\ \hline
7 & 1.24 & 7.0 & 14,2 & 14,2 & 93.5 & 14,1 & 3 & 15,1 & 2.2 &  & 53 & 1.19 & 23.1 & 11,7 & 10,7 & 68.1 & 9,7 & 15.4 & 10,6 & 8.1 \\ \hline
8 & 1.20 & 9.7 & 13,3 & 13,3 & 83.9 & 14,3 & 5.2 & 13,2 & 4.1 &  & 54 & 1.24 & 7.6 & 14,2 & 15,1 & 79.3 & 15,2 & 14.3 & 16,1 & 6 \\ \hline
9 & 1.27 & 30.0 & 9,9 & 9,9 & 75.5 & 10,9 & 14.6 & 10,10 & 8 &  & 55 & 1.22 & 27.7 & 9,8 & 9,8 & 82.3 & 9,9 & 7.1 & 8,8 & 4 \\ \hline
10 & 1.19 & 30.0 & 9,9 & 9,8 & 62.5 & 9,9 & 13.1 & 10,8 & 11.3 &  & 56 & 0.84 & 23.2 & 7,5 & 7,4 & 65.1 & 7,5 & 33.3 & 6,5 & 1.5 \\ \hline
11 & 1.16 & 28.3 & 9,8 & 9,7 & 58.4 & 8,8 & 21.9 & 9,8 & 9.6 &  & 57 & 1.26 & 20.3 & 11,6 & 11,6 & 95.6 & 12,6 & 3.6 & 12,5 & 0.3 \\ \hline
12 & 1.17 & 26.6 & 9,7 & 9,7 & 48.4 & 9,8 & 32.7 & 8,8 & 18.6 &  & 58 & 0.88 & 22.9 & 7,5 & 7,5 & 90.2 & 8,4 & 4.1 & 8,5 & 3 \\ \hline
13 & 1.22 & 23.4 & 11,7 & 10,7 & 77.6 & 9,8 & 10.4 & 10,8 & 5.9 &  & 59 & 1.19 & 27.7 & 9,8 & 9,8 & 72.2 & 8,8 & 25.6 & 9,7 & 1.4 \\ \hline
14 & 1.14 & 24.2 & 9,8 & 8,8 & 41.7 & 9,8 & 33.6 & 9,7 & 24.5 &  & 60 & 1.18 & 29.1 & 9,8 & 9,8 & 52.1 & 9,7 & 24.2 & 8,8 & 23.2 \\ \hline
15 & 1.18 & 27.7 & 9,8 & 9,8 & 90 & 8,8 & 7 & 9,7 & 1.5 &  & 61 & 1.21 & 10.3 & 13,3 & 13,3 & 64.9 & 13,2 & 17.7 & 14,2 & 15 \\ \hline
16 & 1.23 & 18.1 & 13,5 & 13,5 & 63.2 & 12,5 & 27.1 & 12,6 & 5.9 &  & 62 & 0.82 & 1.9 & 9,1 & 9,1 & 98.7 & 8,1 & 0.9 & 9,0 & 0.2 \\ \hline
17 & 1.24 & 3.0 & 15,1 & 14,1 & 53 & 15,0 & 18.2 & 15,1 & 17.8 &  & 63 & 1.03 & 19.8 & 10,5 & 10,5 & 56.1 & 9,5 & 34.4 & 9,6 & 6.6 \\ \hline
18 & 1.16 & 9.7 & 13,3 & 12,3 & 47.3 & 13,2 & 26.7 & 12,4 & 12.3 &  & 64 & 0.92 & 25.5 & 7,5 & 7,5 & 58.6 & 8,5 & 35.6 & 7,6 & 4.7 \\ \hline
19 & 1.20 & 23.1 & 11,7 & 11,7 & 89.2 & 11,6 & 6 & 10.7 & 3.1 &  & 65 & 1.23 & 27.9 & 9,8 & 9,8 & 67.2 & 10,8 & 18.3 & 10,7 & 12.6 \\ \hline
20 & 1.21 & 21.1 & 11,6 & 11,6 & 94.9 & 11,5 & 3.2 & 10,6 & 1.4 &  & 66 & 1.03 & 20.7 & 9,5 & 9,5 & 91.3 & 10,5 & 4.7 & 10,4 & 3 \\ \hline
21 & 1.21 & 25.3 & 8,6 & 8,6 & 95.8 & 7,7 & 2.8 & 7,6 & 1 &  & 67 & 1.19 & 13.6 & 12,4 & 12,4 & 76.5 & 11,4 & 22.1 & 12,3 & 0.8 \\ \hline
22 & 1.18 & 26.6 & 9,8 & 8,8 & 48.4 & 9,7 & 27.4 & 9,8 & 21.8 &  & 68 & 0.84 & 25.5 & 7,5 & 6,5 & 44.6 & 7,5 & 37.9 & 7,4 & 16.6 \\ \hline
23 & 1.14 & 27.8 & 9,8 & 8,8 & 54.5 & 9,8 & 23.6 & 9,7 & 21.3 &  & 69 & 0.87 & 25.5 & 7,5 & 7,5 & 83.1 & 7,4 & 11.2 & 6,5 & 5.3 \\ \hline
24 & 0.94 & 10.9 & 10,3 & 10,3 & 58 & 10,2 & 23.5 & 9,3 & 17.8 &  & 70 & 1.11 & 4.8 & 12,1 & 12,1 & 50.9 & 11,1 & 38.7 & 11,2 & 9.6 \\ \hline
25 & 1.21 & 27.0 & 9,8 & 9,8 & 53.7 & 10,8 & 16.9 & 10,7 & 9.1 &  & 71 & 1.17 & 25.5 & 9,7 & 9,7 & 84.4 & 8,8 & 9.3 & 9,8 & 5.5 \\ \hline
26 & 1.33 & 23.2 & 11,7 & 11,7 & 60.1 & 12,7 & 16.5 & 12,8 & 10 &  & 72 & 0.87 & 25.0 & 7,5 & 7,5 & 51.2 & 7,4 & 45.4 & 6,5 & 3 \\ \hline
27 & 1.29 & 25.7 & 10,8 & 10,8 & 42.4 & 11,7 & 36.3 & 11,8 & 16.5 &  & 73 & 0.97 & 8.2 & 11,1 & 11,1 & 40.9 & 12,1 & 34.1 & 12,0 & 23.7 \\ \hline
28 & 1.15 & 3.7 & 14,1 & 14,0 & 62.9 & 13,1 & 27.8 & 13,0 & 3.4 &  & 74 & 0.93 & 22.9 & 7,5 & 7,5 & 75.7 & 8,5 & 16.7 & 7,6 & 6.1 \\ \hline
29 & 1.22 & 27.7 & 9,8 & 9,8 & 87.6 & 8,8 & 5.5 & 9,9 & 4.7 &  & 75 & 1.27 & 3.7 & 15,1 & 14,1 & 66.3 & 15,1 & 33.5 & 14,2 & 0.1 \\ \hline
30 & 1.18 & 1.3 & 14,1 & 14,1 & 50.9 & 14,0 & 30.5 & 13,1 & 9.9 &  & 76 & 1.37 & 18.0 & 13,6 & 13,6 & 93.5 & 14,6 & 2.8 & 12,6 & 1.3 \\ \hline
31 & 1.12 & 25.3 & 9,7 & 9,7 & 54.1 & 9,6 & 21.8 & 8,7 & 20.6 &  & 77 & 1.15 & 25.5 & 9,7 & 9,7 & 80.4 & 10,7 & 8.2 & 10,6 & 7 \\ \hline
32 & 1.18 & 27.0 & 9,8 & 9,8 & 73.2 & 8,8 & 17.3 & 9,7 & 8.2 &  & 78 & 1.18 & 16.1 & 11,5 & 11,5 & 73.4 & 12,4 & 14.6 & 12,5 & 10 \\ \hline
33 & 1.24 & 25.5 & 10,8 & 10,8 & 79.5 & 9,9 & 7.6 & 10,7 & 6.3 &  & 79 & 1.44 & 27.7 & 11,10 & 10,10 & 39.8 & 11,9 & 31.7 & 11,10 & 12.2 \\ \hline
34 & 1.29 & 25.5 & 10,8 & 10,8 & 82.5 & 11,8 & 14.8 & 10,9 & 1.2 &  & 80 & 1.31 & 12.6 & 14,4 & 14,4 & 64.4 & 13,4 & 23.3 & 14,3 & 7.2 \\ \hline
35 & 1.22 & 18.6 & 12,5 & 12,5 & 73.9 & 13,5 & 9.6 & 12,6 & 7.4 &  & 81 & 1.09 & 4.0 & 13,1 & 13,0 & 64.6 & 13,1 & 19.9 & 12,1 & 9.9 \\ \hline
36 & 1.18 & 27.7 & 9,8 & 8,8 & 40.2 & 9,7 & 35 & 9,8 & 18.6 &  & 82 & 1.22 & 23.1 & 10,7 & 10,7 & 42.6 & 9,7 & 36.5 & 10,6 & 15.5 \\ \hline
37 & 1.13 & 27.7 & 9,8 & 8,8 & 53.1 & 9,7 & 42.4 & 8,7 & 3.3 &  & 83 & 1.05 & 26.5 & 8,7 & 8,7 & 96.5 & 7,7 & 1.7 & 9,6 & 0.7 \\ \hline
38 & 1.17 & 20.7 & 10,6 & 10,6 & 58.9 & 11,6 & 24.2 & 11,5 & 14.2 &  & 84 & 1.12 & 12.2 & 12,3 & 12,3 & 76.7 & 11,3 & 10.7 & 11,4 & 10.1 \\ \hline
39 & 1.18 & 22.0 & 11,6 & 10,6 & 66.5 & 11,6 & 22.5 & 11,5 & 9.9 &  & 85 & 1.82 & 13.5 & 19,6 & 19,6 & 56 & 18,6 & 21 & 19,5 & 8.5 \\ \hline
40 & 1.21 & 16.1 & 12,5 & 12,5 & 53.3 & 12,4 & 18.5 & 11,5 & 18 &  & 86 & 1.31 & 6.0 & 15,2 & 15,2 & 56.1 & 14,2 & 31.6 & 15,1 & 5.6 \\ \hline
41 & 0.99 & 0.0 & 12,0 & 12,0 & 97.2 & 13,0 & 2.8 & 12,1 & 0 &  & 87 & 1.89 & 19.5 & 18,9 & 17,9 & 61.9 & 18,9 & 12.2 & 17,8 & 9.3 \\ \hline
42 & 1.27 & 25.5 & 10,8 & 9,8 & 26.6 & 10,8 & 21.9 & 10,7 & 19.8 &  & 88 & 1.32 & 24.8 & 11,8 & 10,8 & 47.7 & 11,8 & 27.1 & 11,7 & 22.9 \\ \hline
43 & 1.23 & 30.0 & 9,9 & 9,8 & 68.1 & 9,9 & 30.9 & 10,8 & 0.5 &  & 89 & 0.95 & 16.1 & 10,3 & 10,3 & 55.5 & 10,2 & 23.6 & 11,2 & 17.1 \\ \hline
44 & 1.18 & 13.5 & 12,4 & 12,4 & 88.2 & 13,3 & 7.8 & 13,4 & 2.2 &  & 90 & 1.70 & 18.8 & 16,8 & 15,8 & 50.6 & 15,7 & 12.1 & 16,7 & 11.3 \\ \hline
45 & 1.18 & 27.7 & 9,8 & 9,8 & 74.5 & 8,8 & 10.5 & 9,9 & 7.4 &  & 91 & 0.95 & 6.6 & 10,2 & 10,2 & 50.9 & 11,2 & 44.6 & 11,1 & 3.8 \\ \hline
46 & 1.17 & 10.5 & 13,3 & 13,3 & 97.6 & 12,3 & 1.5 & 14,3 & 0.8 &  &  \\ \hline
\end{tabular}
\end{table*}

\section{Examples of successfully classified HRTEM images of CNTs}
\label{sec:exp_images_success}

\begin{figure}[htb]
    \includegraphics[width=3.2in]{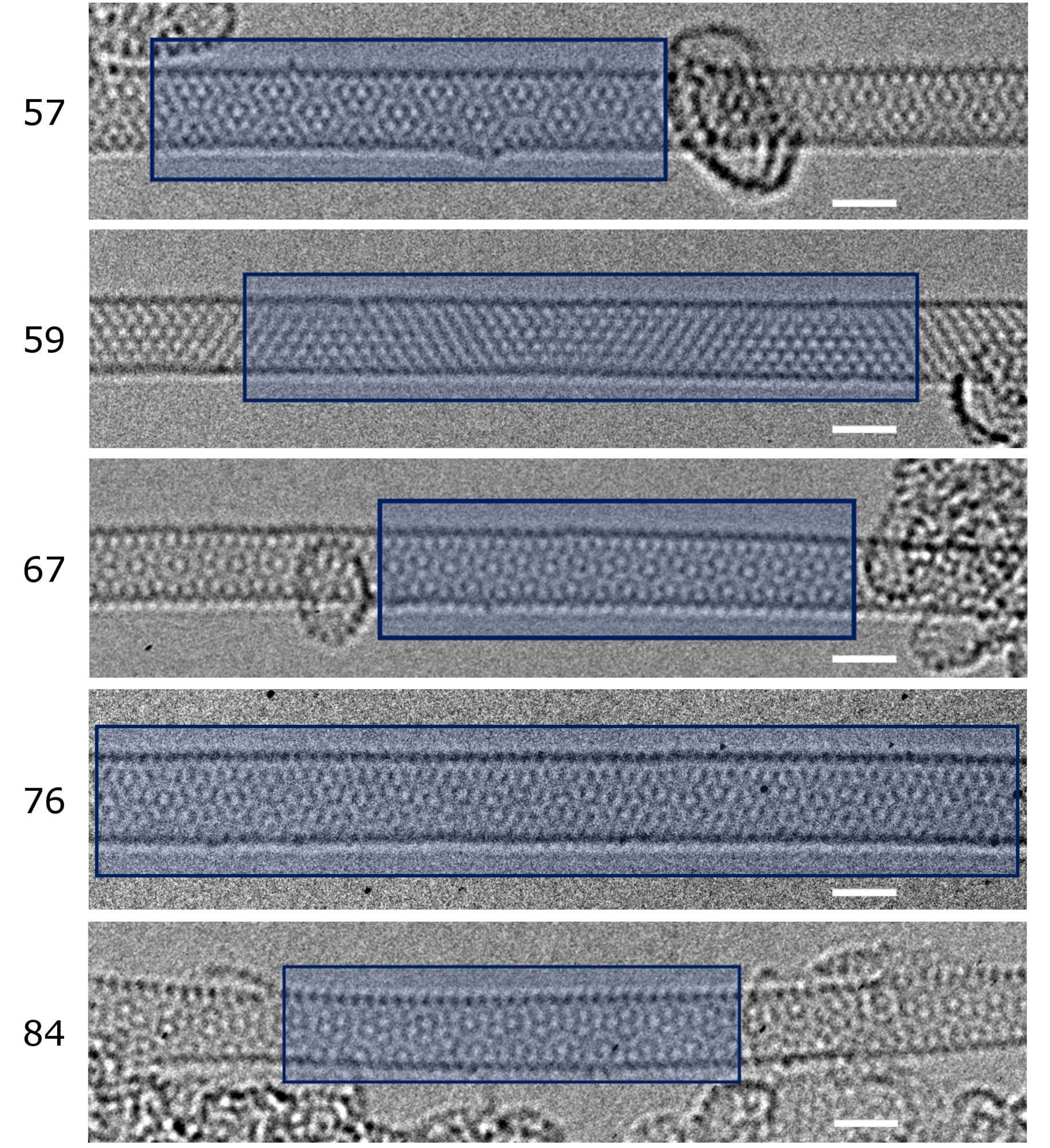}
    \caption{(Color online) Five examples of HRTEM images of CNTs where the same chiral indecis were obtained with the manual and deep learning methods : These are (11,6), (9,8), (12,4), (13,6), and (12,3). Scale bars are 1~nm in all the micrographs. The section analyzed by the deep learning tool is highlighted in blue. The numbers correspond to the entries in Tab~\ref{tab:comparison}.}
    \label{fig:exp_images_success}
\end{figure}

In about 70\% of the cases, the conventional and deep learning methods yield the same results in the case of our experimental data set of 91 images (see Tab~\ref{tab:comparison}). Figure~\ref{fig:exp_images_success} shows some examples where both methods resulted in the same chiral indecis. The images are of high quality and the moiré created by the superposition of the front and back wall of the CNTs is clearly visible. Hoover, the CNTs contain some defects which does not negatively affect the automatic classification in these cases indicating the robustness of our approach.

\section{Classification errors with experimental HRTEM images of CNTs}
\label{sec:error_case_exp}

\begin{figure}[htb]
    \includegraphics[width=3.2in]{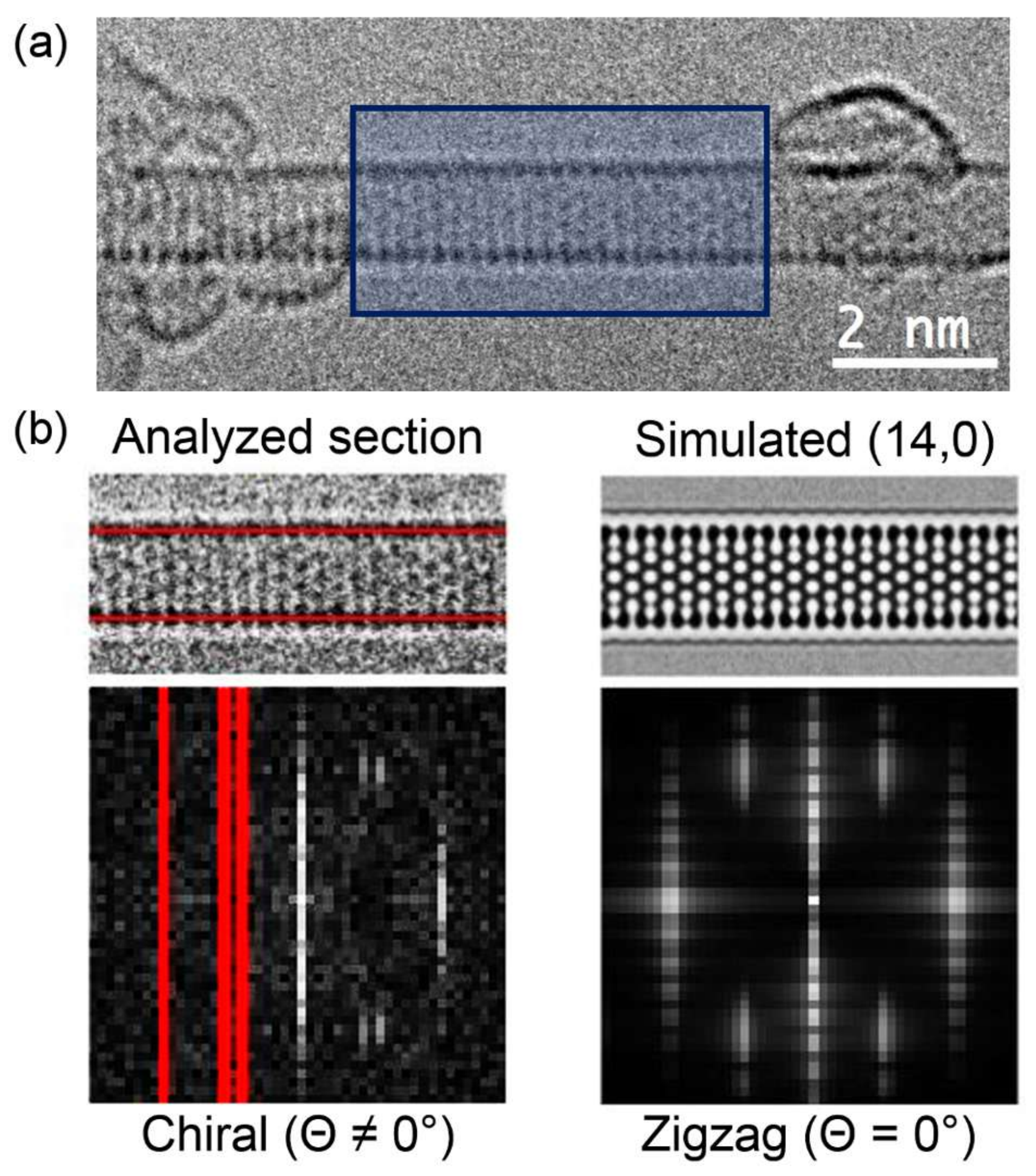}
    \caption{(Color online)
Example of an error case: (a) Experimental HRTEM image of the analyzed CNT (the section analyzed by the deep learning tool is highlighted in blue). The deep learning tool classifies the chirality as (14,1), see entry 28 in Tab~\ref{tab:comparison}. (b) comparison of the FT of the analyzed section of the experimental image (left) showing sets of layer lines of a chiral nanotube (non-zero chiral angle), and of a simulated image of a (14,0) zigzag CNT.}
    \label{fig:error_case_exp}
\end{figure}

A disagreement between the conventional and deep learning-based analysis techniques means that one of the two, or both gave a wrong result. In some cases it is not trivial to determine which technique led to a false classification, if not both. An example of disagreement between the two techniques is given in Fig.~\ref{fig:error_case_exp} (CNT 28 in Tab~\ref{tab:comparison}). Here, the automatic technique leads to the classification of the observed nanotube as a (14,0), a zigzag nanotube with a 0 $^{\circ}$ chiral angle. In this case, the result is easily contradicted when looking at the FT of the HRTEM image, which unambiguously shows three sets of layer lines, indicative of a chiral nanotube. Verifying the results obtained by the deep learning tool by a systematic check of the FT of the HRTEM image can therefore be very helpful. In particular, refuting the hypothesis of an armchair CNT is straightforward with this method. Other cases, where it is not trivial to determine which technique led to a wrong classification also exist. Therefore, it should be noted that high-quality HRTEM images with clean nanotubes, suspended over sufficient lengths, and with low defect density minimize the uncertainties in chirality determination with both methods.

\section{Performance of the classification system with simulated images outside the range of the training data set}
\label{sec:out_of_range_performance}

\begin{figure*}[htb]
    \includegraphics[width=6.4in]{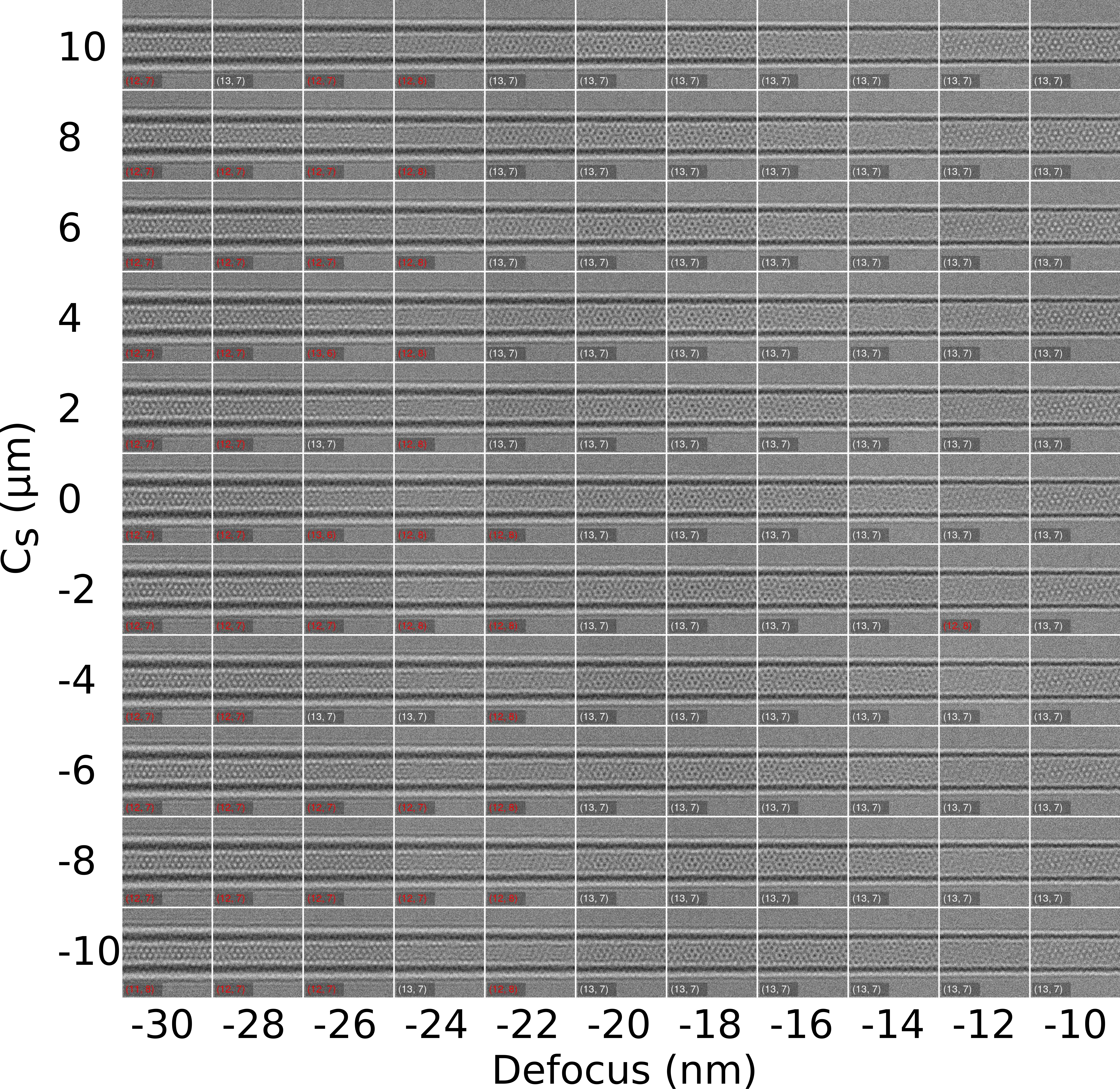}
    \caption{(Color online) Simulated HRTEM image of a (13,7) CNT as a function of defocus and spherical aberration. All other aberration coefficients are set to zero. The classification result with the deep learning tool is indicated on each panel.}
    \label{fig:defocus_cs_a}
\end{figure*}

\begin{figure*}[htb]
    \includegraphics[width=6.4in]{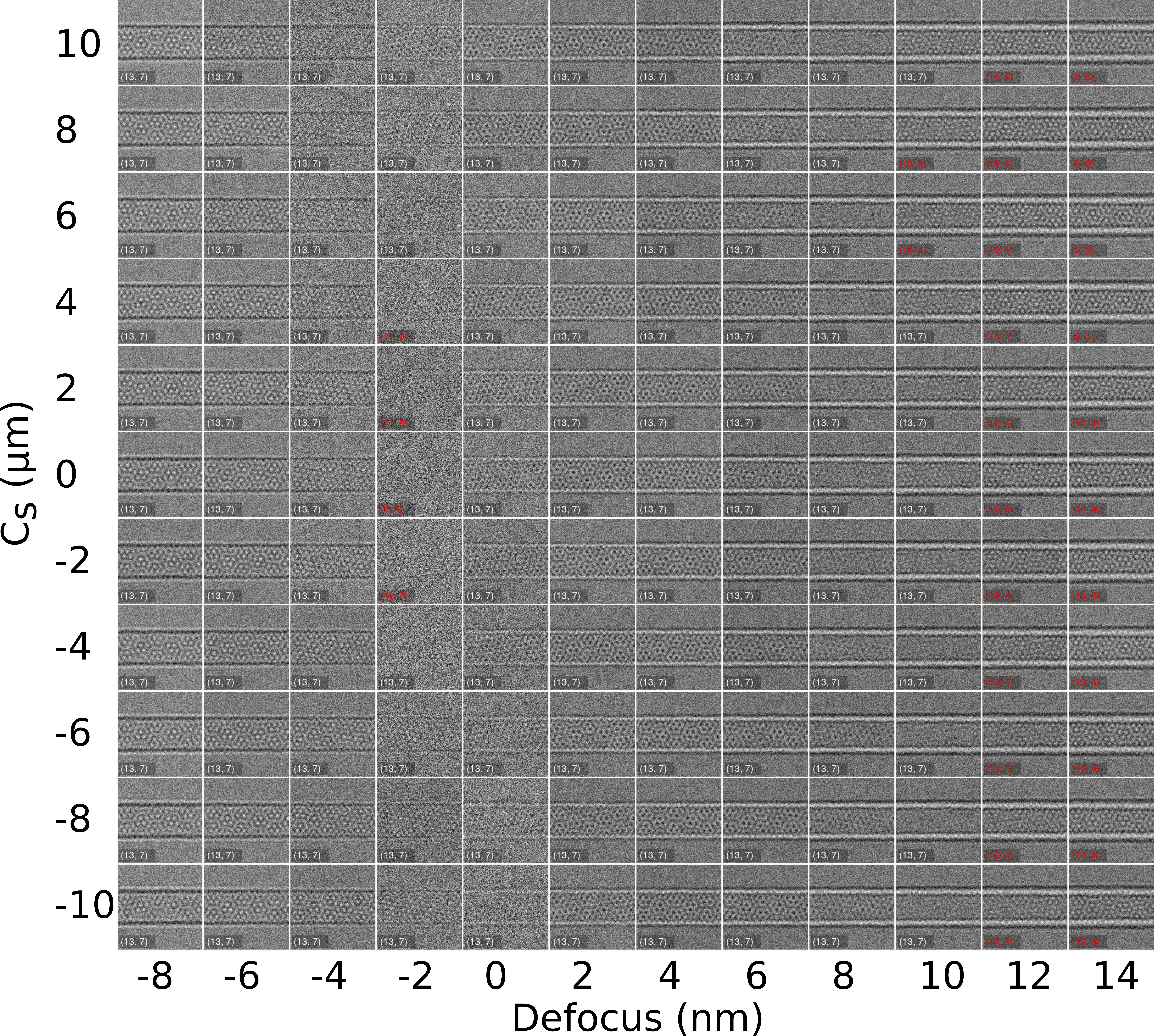}
    \caption{(Color online) Same as Fig.~\ref{fig:defocus_cs_a}, with defocus varied between -8 and +14~nm.}
    \label{fig:defocus_cs_b}
\end{figure*}

\begin{figure*}[htb]
    \includegraphics[width=6.4in]{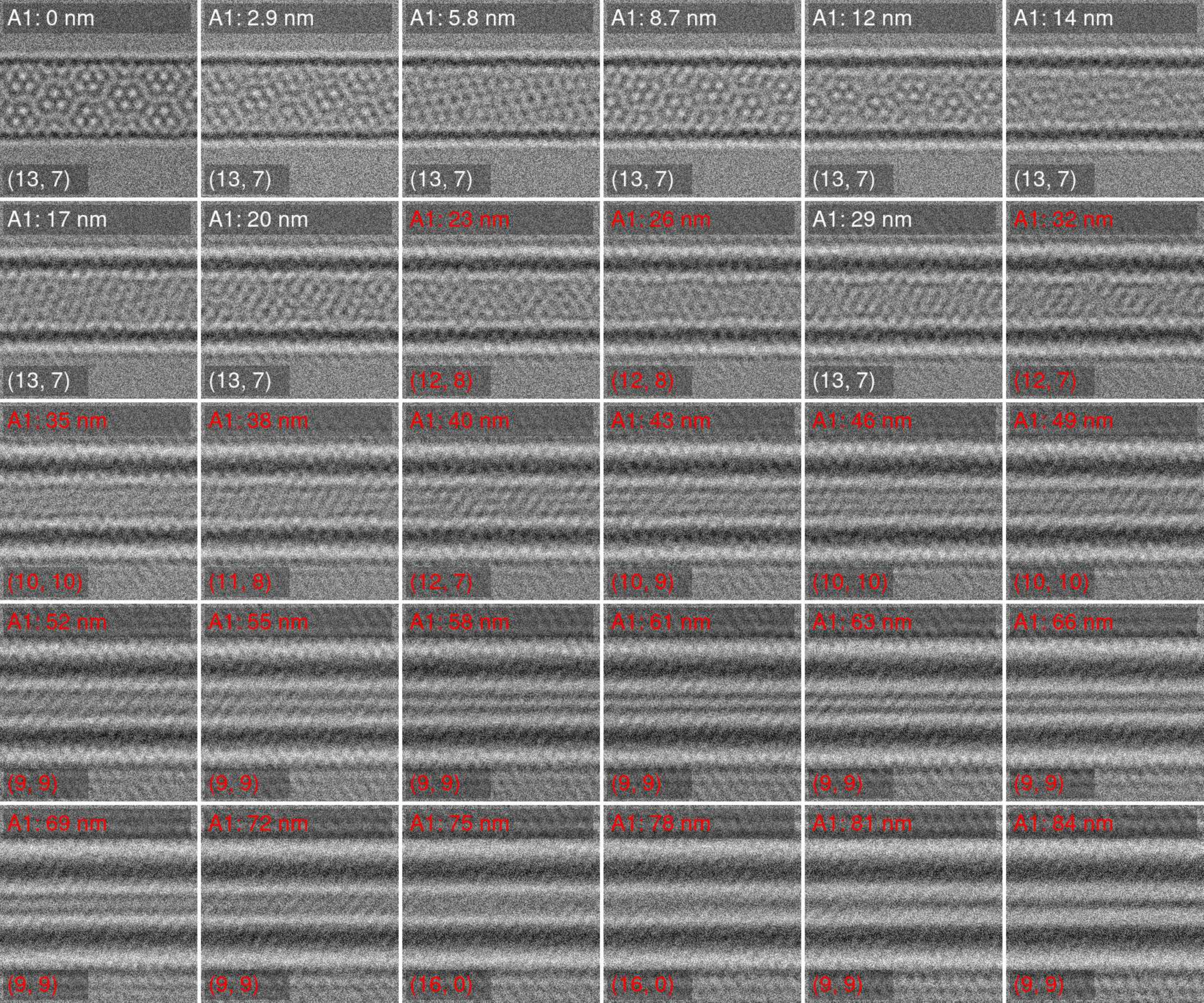}
    \caption{(Color online) Simulated HRTEM image of a (13,7) CNT as a function of twofold astigmatism oriented 30$^\circ$ with respect to the CNT axis. Defocus and spherical aberration are set to -7~nm and 1~$\mu$m, respectively. All other aberration coefficients are set to zero. The classification result with the deep learning tool is indicated on each panel.}
    \label{fig:twofold_astigmatism_30}
\end{figure*}

In order to account for possible situations where experimental images are obtained with aberration coefficients outside the ranges of Tab.~1 (main text), we have evaluated the performance of our automatic classification system with images simulated within much larger ranges of defocus, spherical aberration and two-fold astigmatism values. As illustrated in Figs~\ref{fig:defocus_cs_a} and~\ref{fig:defocus_cs_b}, aside from peculiar cases where image contrast almost vanishes, our deep learning approach successfully classifies the chirality of CNTs if the defocus and the spherical aberration stay between -20 nm to +8 nm and -10 $\mu m$ to +10 $\mu m$, respectively. Similarly, values of up to 20 nm of two-fold astigmatism do not jeopardize the chirality assignment (see Fig.~\ref{fig:twofold_astigmatism_30}). This extended range of aberration variations is more in line with the experimental variability of HRTEM and we thus assume that the optical parameters of the experimental images are not the main source of misclassifications. Moreover, we noticed that the cases where the manual and automatic procedures are in disagreement do not correspond to images with visible effects of aberrations.

\bibliography{bibliography}